\documentstyle[multicol,aps,epsfig]{revtex}
\begin{document}

\draft
\title{
Antiferromagnetism and d-wave superconductivity in cuprates: a
cluster DMFT study.
}
\author{
A.I. Lichtenstein$^1$ and M.I. Katsnelson$^2$
}
\address{
$^1$ University of Nijmegen, 6525 ED Nijmegen, The Netherlands \\
$^2$ Institute of Metal Physics, 620219 Ekaterinburg , Russia 
}
\maketitle
\begin{abstract}
We present a new approach to investigate the coexistence of antiferromagnetism
and d-wave superconductivity in the two dimensional extended Hubbard model
within a numerically exact cluster dynamical mean-field approximation.
Self-consistent solutions with two non-zero order parameters exists in
the wide range of doping level and temperatures. 
A linearized equation for energy spectrum near the Fermi level have been solved.
The resulting d-wave gap has the correct magnitude and k-dependence
but some distortion compare to the pure  
d$_{x^2-y^2}$ superconducting order parameter due to the presence of
underlying antiferromagnetic ordering.
\end{abstract}

\begin{multicols}{2}

A microscopic theory of High-temperature superconducting cuprates (HTSC)
is still far from the final understanding\cite{PWA,Scalapino,Pines}. One of
the most important recent experimental achievements was the discovery of pseudo-gap (PG)
phenomenon above superconducting transition temperatures (T$_{c}$)\cite{PGAP}
and existence of a sharp 41-meV resonance below T$_{c}$ related with some
collective antiferromagnetic excitations\cite{Neutr}. Recent neutron
scattering experiments\cite{Keimer} provide a new insight for the interesting
problem on the origin of a condensation energy. 
The coexistence of an antiferromagnetism
(AFM) and d-wave superconductivity (d-SC) in cuprate could be a natural
way of discussing such different HTSC phenomena.
This require a quantitative electronic structure theory including two 
different type of the order parameters: AFM and d-SC. Within such approach
one can in principle analyzed the phase diagram of HTSC cuprate and
interplay between antiferromagnetism and d-wave superconductivity\cite{SO5}.

A minimal theoretical tool for cuprates consists of the two-dimensional
Hubbard model\cite{PWA}. The importance of including the realistic
tight-binding spectrum obtained from the LDA-band structure analyses\cite
{OKA} was realized during the last years. Unfortunately, a most accurate
Quantum Monte-Carlo (QMC) simulation of hole-doped 2d-Hubbard model has a
difficulty to describe an interesting part of
the HTSC phase diagram near 15\% doping at
the low temperature due to so-called sign-problem\cite{White}. The
perturbation theory of d-SC\cite{Bickers} 
ignore the vertex corrections in the strong correlation case of HTSC.
Great progress in the theory of the interacting fermions results from 
the developing of the dynamical mean-field theory\cite{Metzner,DMFT}.
While the antiferromagnetic phase is easy to incorporate in the single-impurity
DMFT-approach\cite{DMFT}, the d-wave superconductivity requires a cluster
generalization of DMFT.
Different cluster-DMFT scheme have been proposed\cite{DMFT,DCA}
and the recent application to the problem of the pseudogap in HTSC\cite{DCAPG}
have shown the efficiency of the cluster DMFT approach. The investigation of
paramagnetic phase of two dimensional Hubbard model could be
simplified using a translational symmetry\cite{DCA},
while the problem of a coexistence
of AFM and dSC demands a broken-symmetry cluster calculation. This is equivalent    
to multi-orbital DMFT approach\cite{LDAPP} and could be solved with
QMC method\cite{Rozenberg}. 

In this Letter we investigate the problem of antiferromagnetism and d-wave
superconductivity in two-dimensional Hubbard model within a cluster DMFT
scheme. 

\begin{figure}[t!]
\begin{center}
\vskip  0.5cm
\epsfig{figure=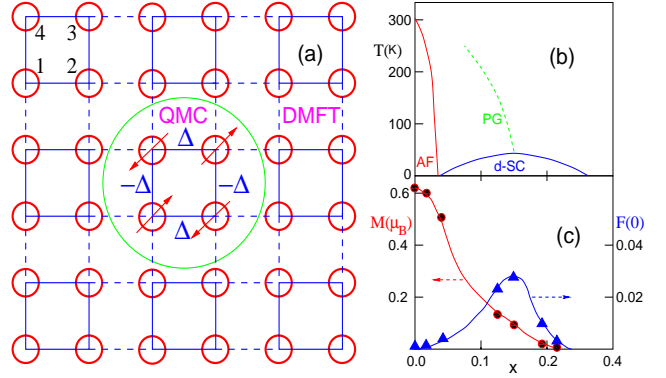,width=8.5cm,height=5.0cm,angle=0}
\vskip  0.4cm
\end{center}
\caption{
(a) Schematic representation of antiferromagnetic d-wave 2x2 periodically repeated cluster;
(b) generic phase diagram of HTSC materials;
(c) The calculated values of two order parameters 
(local magnetic moment and d-SC equal time Green function F$^{01}$) for different  hole doping (x)
at the inverse temperature $\beta=60$ eV$^{-1} (T=190 K)$.
}
\label{AFMSC}
\end{figure}

The minimal cluster which allow us to study both AFM and d-SC order
parameters on the equal footing consists of 2x2 system in the effective
DMFT-medium.
We start with the extended Hubbard model on the square lattice:

\[
H=\sum_{ij}t_{ij}c_{i\sigma }^{+}c_{j\sigma }+\sum_{i}U_{i}n_{i\uparrow
}n_{i\downarrow }
\]
where $t_{ij}$ is an effective hopping and $U_{i}$ local Coulomb
interactions. We chose nearest-neighbor hopping $t=0.25$ eV and the next
nearest hopping $t^{\prime }/t=-0.15$ for the model of La$_{2-x}$Sr$_{x}$CuO$%
_{4}$\cite{OKA}. The total band width is W=8t and the Coulomb interactions
set to be U/W=0.6. Let us introduce the ''super-site'' as an 2x2 square (Fig.\ref{AFMSC}).
The numeration of the atoms in the super-site is also shown in the Fig.\ref{AFMSC}.
It is useful to introduce the superspinor $C_{i}^{+}=\{c_{i\alpha }^{+}\}
$ where $\alpha =1,2,3,4.$ (the spin indices are not shown). Taking into
account the spin degrees of freedom, this is 8-component superspinor creation operator.
Then the crystal Green function for the Hubbard model can be rewritten as 
\[
G\left( {\bf k,}i\omega \right) =\left[ i\omega +\mu -h\left( {\bf k,}%
i\omega \right) \right] ^{-1}
\]
where $h\left( {\bf k,}i\omega \right) $ is the effective hopping
supermatrix and $\mu $ is the chemical potential. 
For simplicity we will write all the formulas in the
nearest-neighbor approximations: \ 
\begin{equation}
h\left( {\bf k,}i\omega \right) =\left( 
\begin{array}{cccc}
\Sigma _{0} & t_{x}K^+_x & 0 & t_{y}K^+_y \\ 
t_{x}^{\ast }K^-_x & \Sigma _{0} & t_{y}K^+_y & 0 \\ 
0 & t_{y}^{\ast }K^-_x & \Sigma _{0} & t_{x}^{\ast}K^-_x \\ 
t_{y}^{\ast }K^-_y & 0 & t_{x}K^+_x & \Sigma _{0}
\end{array}
\label{hkom}
\right) 
\end{equation}
where $K^{\pm}_{x(y)}=1+\exp\left(\pm ik_{x(y)}\right)$ and 
each element is 2$\times $2  matrix in the spin space. 
Within cluster-DMFT
approach we introduce intraatomic self-energy $\Sigma _{0}$ and
interatomic self-energies $\Sigma _{x},$ $\Sigma _{y}$ and the both are
intra-site in the sense of the super-site: 
\[
\Sigma \left( i\omega \right) =\left( 
\begin{array}{cccc}
\Sigma _{0} & \Sigma _{x} & 0 & \Sigma _{y} \\ 
\Sigma _{x}^{\ast } & \Sigma _{0} & \Sigma _{y} & 0 \\ 
0 & \Sigma _{y}^{\ast } & \Sigma _{0} & \Sigma _{x}^{\ast } \\ 
\Sigma _{y}^{\ast } & 0 & \Sigma _{x} & \Sigma _{0}
\end{array}
\right) 
\]
The effective Hamiltonian defined
through the translationaly invariant self-energy corresponds to $t_{x}=$ $t+\Sigma
_{x},$ $t_{y}=t+\Sigma _{y}$ renormalized energy dependent hopping. The
functions $\Sigma _{0}\left( i\omega \right) ,$ $\Sigma _{x}\left( i\omega
\right) ,$ $\Sigma _{y}\left( i\omega \right) $ are found self-consistently
within DNFT scheme\cite{DMFT} and for the
d-wave superconduction state $\Sigma _{x}\neq \Sigma _{y}.$
It is straightforward to generalize this scheme for next-nearest neighbor
(or more extended) hopping as well as the long-range Green function and self-energy.
In this case we can renormalized also the second-nearest
hopping: $t_{xy}=t'+\Sigma _{xy}$ for the 2x2 cluster, where $\Sigma _{xy}$ (or $\Sigma^{02}$) is
the non-local self-energy in $xy$ direction. 

According to the prescription of DMFT scheme\cite{DMFT}, we can write the matrix
equation for the so-called bath Green function matrix ${\cal G}$
which account for a double counting correction for the self-energy: 
\[
{\cal G}^{-1}\left( i\omega \right) =G^{-1}\left( i\omega \right) +\Sigma
\left( i\omega \right) 
\]
where the local cluster Green function matrix is equal to $G_{\alpha \beta
}\left( i\omega \right) =$ $\sum\limits_{{\bf k}}G_{\alpha \beta }\left( 
{\bf k,}i\omega \right)$ , 
and summation is run over the Brillouin zone of the square lattice. 
Note that in the Eq.(\ref{hkom}) we use  translationaly invariant
self-energy obtained from the cluster DMFT similar to the dynamical cluster approximation\cite{DCA}. 
The present ``matrix'' form of a cluster
DMFT with the self-energy which is not periodic inside the cluster 
allow us to study a multicomponent ordered state.

In this case we have the standard DMFT problem with four ``orbital'' states
per super-site. It is solved by multi-orbital QMC technique\cite{Rozenberg}. 
For the problem of a coexistence of magnetic ordering and superconductivity 
one can use the generalized Nambu technique\cite{IKS}. We introduce the superspinor 
\[
\Psi _{i}^{+}(\tau )\equiv (\psi _{1i}^{+},\psi _{2i}^{+},\psi
_{3i}^{+},\psi _{4i}^{+})=\left( c_{i\uparrow }^{+},c_{i\downarrow
}^{+},c_{i\uparrow },c_{i\downarrow }\right) 
\]
and the anomalous averages describing the (collinear) antiferromagnetism $%
\left\langle c_{i\uparrow }^{+}c_{j\downarrow }\right\rangle $ and the
superconductivity $\Delta _{ij}=$ $\left\langle c_{i\downarrow }c_{j\uparrow
}\right\rangle $.

We use the generalization of the Hirsch-Fye QMC-algorithm\cite{Hirsch} for
superconducting problem\cite{Georges}. In  the 4-spinor case a 
discrete Hubbard-Stratonovich transformation has the following form:

\begin{eqnarray}
\exp [-\Delta \tau U_{i}n_{i\uparrow }n_{i\downarrow }+\frac{\Delta \tau
U_{i}}{2}(n_{i\uparrow }+n_{i\downarrow })]=   \nonumber \\ 
\frac{1}{2}\sum_{\sigma = \pm 1}\exp [\lambda _{i}\sigma 
(\psi _{1i}^{+}\psi _{1i}-\psi _{2i}^{+}\psi
_{2i}-\psi _{3i}^{+}\psi _{3i}+\psi _{4i}^{+}\psi _{4i})]  \nonumber 
\end{eqnarray}
where $\lambda _{i}=\frac{1}{2}{\rm arccosh}[\exp (\frac{1}{2}\Delta \tau
U_{i})] $.

We will take into account only the singlet pairing and in the case of  d-wave there are following
nonzero elements of $\Delta $ matrix: 
$\Delta =\Delta _{12}=-\Delta _{23}=\Delta _{34}=-\Delta _{41}$. 
One can chose $\Delta _{ij}$ to be real and therefore symmetric: $\Delta
_{ij}=\Delta _{ji}$. Separating normal and anomalous parts of the Green
function we have 
\[
G\left( {\bf k,}\tau ,\tau ^{\prime }\right) =\left( 
\begin{array}{cc}
G\left( {\bf k,}\tau ,\tau ^{\prime }\right) & F\left( {\bf k,}\tau
,\tau ^{\prime }\right) \\ 
F^{+}\left( {\bf k,}\tau ,\tau ^{\prime }\right) & -G\left( -{\bf k,}%
\tau ^{\prime },\tau \right)
\end{array}
\right) 
\]
where $G\left( {\bf k,}\tau ,\tau ^{\prime }\right) =-\left\langle
T_{\tau }C_{{\bf k}}\left( \tau \right) C_{{\bf k}}^{+}\left( \tau ^{\prime
}\right) \right\rangle ,$ $F\left( {\bf k,}\tau ,\tau ^{\prime }\right)
=-\left\langle T_{\tau }C_{{\bf k}}\left( \tau \right) C_{-{\bf k}}\left(
\tau ^{\prime }\right) \right\rangle $ are the matrices in spin and
``orbital'' space. It is convenient to expand the anomalous Green function
in Pauli matrices 
$F=\left( F^{0}+{\bf F\sigma }\right) i\sigma ^{y}$ 
and use the symmetry properties \cite{GV}: 
\begin{eqnarray*}
F^{0}\left( {\bf k,}\tau ,\tau ^{\prime }\right) &=&F^{0}\left( -{\bf k,}%
\tau ^{\prime },\tau \right) \\
{\bf F\left( k,\tau ,\tau ^{\prime }\right) } &&{\bf =-F}\left( -{\bf k,}%
\tau ^{\prime },\tau \right)
\end{eqnarray*}
Therefore in the collinear antiferromagnetic case with d-wave superconductivity the $%
4\times 4$ spinor formalism is reduced to $2\times 2$ one with the following
spin-matrix form of the local Green function for our super-site:

\[
G\left( \tau ,\tau ^{\prime }\right) =\left( 
\begin{array}{cc}
G_{\uparrow }\left( \tau ,\tau ^{\prime }\right) & F\left( \tau ,\tau
^{\prime }\right) \\ 
F(\tau ,\tau ^{\prime }) & -G_{\downarrow }\left( \tau ^{\prime },\tau
\right)
\end{array}
\right) 
\]
and the QMC formalism for the antiferromagnetic superconducting state is
equivalent to the previous non-magnetic one\cite{Georges}. Using the
discretization of $[0,\beta ]$ interval with $L$-time slices ($\Delta \tau
=\beta /L$ and $\beta =1/T$ the inverse temperature) $G_{\sigma }$ and $F$
Greens functions are matrices of the dimension $2NL$, where $N$ is the
number of atoms in the cluster.
After a Fourier transform to the Matsubara frequencies we have:

\[
G\left( i\omega \right) =\left( 
\begin{array}{cc}
G_{\uparrow }\left( i\omega \right) & F\left( i\omega \right) \\ 
F(i\omega ) & -G_{\downarrow }^{\ast }\left( i\omega \right)
\end{array}
\right) 
\]
In superconducting states\cite{DMFT} the self-energy defined as:

\[
{\cal G}^{-1}\left( i\omega \right) -G^{-1}\left( i\omega \right) =\left( 
\begin{array}{cc}
\Sigma _{\uparrow }\left( i\omega \right) & S\left( i\omega \right) \\ 
S(i\omega ) & -\Sigma _{\downarrow }^{\ast }\left( i\omega \right)
\end{array}
\right), 
\]
and the crystal Green functions is equal to:

\[
G^{-1}\left( {\bf k,}i\omega \right) =\left( 
\begin{array}{cc}
i\omega +\mu -h\left( {\bf k,}i\omega \right)  & s\left( {\bf k,}i\omega
\right)  \\ 
s\left( {\bf k,}i\omega \right)  & i\omega -\mu +h^{\ast }\left( {\bf k,}%
i\omega \right) 
\end{array}
\right) 
\]
where $s\left( {\bf k,}i\omega \right) $ translationaly invariant anomalous
part of self-energy $S(i\omega )$ similar to Eq.(\ref{hkom}).

\begin{figure}[t!]
\begin{center}
\vskip -0.5cm
\epsfig{figure=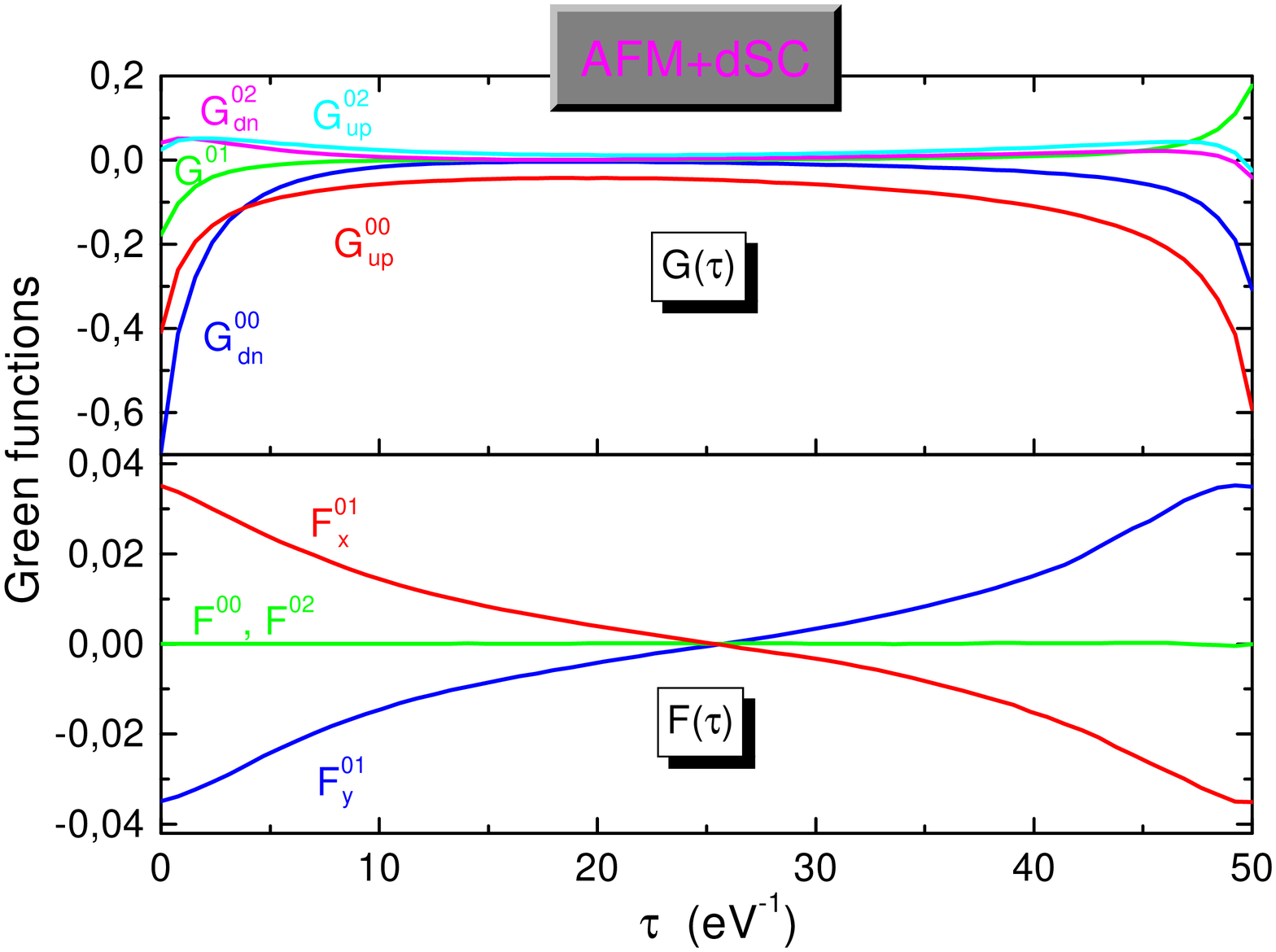,width=9.5cm,height=7.5cm,angle=0}
\vskip -0.4cm
\end{center}
\caption{
Imaginary time normal (G) and superconducting (F) Green functions for the
2x2 cluster DMFT solution  with second-nearest neighbor hopping 
and inverse temperature $\beta=50$ (eV$^{-1}$)
(T=230 K).
}
\label{GTAU}
\end{figure}

The two-component order parameters state which consist of Neel antiferromagnetic state and
d-wave superconducting (Fig.1a) lowered the symmetry of effective cluster-DMFT
problem and did not allowed to use single-atom translational symmetry. The self-consistent
DMFT 8x8 matrix cluster problem with AFM and d-SC general order parameters have been 
solved within QMC scheme with L=64 time slices.
The resulting two order parameters for $\beta=60$ (or T=190 K) and $t'=0$  
was presented in Fig.1c together with the generic HTSC phase
diagram (Fig.1b) as function of the hole doping. In this case the ordered magnetic
moment is directly related with imaginary-time Green function $G_{\sigma}(\tau)$:
$M=G^{00}_{\uparrow }\left(0 \right)-G^{00}_{\downarrow }\left(0 \right)$ 
and for d-SC order parameter we chose a positive value of superconducting 
imaginary time Green function 
$F^{01}\left(0 \right)$. It is interesting that the AFM cluster-DMFT
solution exist for a much higher doping concentration then experimental AFM ordered state
and describe a dynamical mean-field version of AFM-spin fluctuations related
to pseudogap phenomena (PG-region on the fig.1b). The maximum of d-SC order parameter
corresponds to the doping level of about 15$\%$ and agree well with the generic
HTSC phase diagram. The d-SC order parameter is zero close to the
undoped region (x=0), due to presence of a large AFM-gap. When the AFM-gap is closed (x$\sim 5\% $)
the d-SC states developed and after x$\sim 20\% $ it decreasing again since AFM spin-fluctuations
around ($\pi, \pi$) point disappear\cite{Scalapino}. 
 The precise characteristic of the phase diagram including    
the interactions between AFM and d-SC order parameters demands an extensive
cluster-DMFT calculations for different temperatures and doping. 

We would like to note that the existing of d-SC cluster-DMFT solution for
such high temperature does not necessary means that the superconducting transition
temperature is even larger then 190K in our model. A crude estimation shows that
d-SC solution disappear about 300 K for x=0.15 and AFM solutions for x=0 become
unstable at the temperature just above 1000 K. This could be the sign of a ``local''
AFM solutions  or local d-wave solutions.
like a local moment in magnetic systems.
We plane to estimate a proper superconductiong transition temperature in the
future publication.
It is important, that we find no serious sign-problem for all QMC calculations
with various doping level, probably due to ``stabilized'' antiferromagnetic
dynamical mean fields acting on the neighbor atoms to our 2x2 cluster.

\begin{figure}[t!]
\begin{center}
\vskip -0.2cm
\epsfig{figure=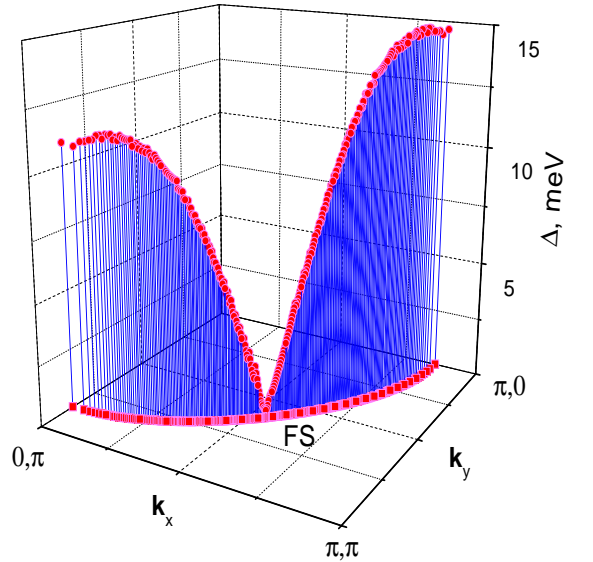,width=10.cm,height=8.0cm,angle=0}
\vskip -0.2cm
\end{center}
\caption{
The d-wave gap function at the Fermi surface 
for x=0.15, $t'=0$ and $\beta=50$ (eV$^{-1}$).
}
\label{GAP}
\end{figure}

The general 
role of next-nearest hopping is the lowering of the van-Hove singularity\cite{OKA}
which increase the density of state at the Fermi level for the doped case and
favored d-SC solution.
There is also change in the spin-fluctuation spectrum related with broadening
of AFM-peak near ($\pi, \pi$) point due to formation of so-called extended
van-Hove singularities with increasing of the $t'$.  
On the Fig.\ref{GTAU} we show one of the AFM-dSC solution with the next 
nearest-neighbor hopping for the 10\% doping level and $\beta=50$ (ev$^-1$).
Resulting local magnetic moment is M=0.28 $\mu_B$ and F(0)=0.036.One could see that
the superconduction order parameter is really of the d$_{x^2-y^2}$ symmetry since
diagonal elements (F$^{00}$) as well as the nearest-neighbors elements (F$^{00}$) 
are all equal to zero and the only nearest neighbor superconducting Green function
(F$^{01}$) is non-zero and moreover change the sign for F$_x$ and F$_y$ components.
The normal local Green function (G$^{00}$) (ploted for the spin-up atom in the Fig.\ref{GTAU})
as well as (G$^{02}$) is spin-split, while
nearest-neighbor Green function (G$^{01}$) has no spin-splitting due to AFM spin
symmetry (see Fig.\ref{AFMSC}).

In order to find the superconducting energy gap we solved the linearized
equation for energy spectrum, assuming that the characteristic energy scale
of $\Sigma _{\sigma }\left( i\omega \right) $ and $S\left( i\omega \right) $
is larger than the SC-gap $\Delta \left( i\omega \right)$. In this case we
could perform analytical energy continuations and the generalized equation
for the energy spectrum has the simple form: 

\[
\det (H-EO)=0
\]
where $H=t\left( {\bf k}\right) +\Sigma (0)-\mu $ , $O=1-\Sigma ^{\prime }(0)
$ and $\Sigma (0)=\int_{0}^{\beta }\Sigma (\tau )d\tau $ , $\Sigma ^{\prime
}(0)=\int_{0}^{\beta }\tau \Sigma (\tau )d\tau $. Note that $\Sigma (0)$ and 
$\Sigma ^{\prime }(0)$ in this expression should be also translationaly invariant
similar to Eq.(\ref{hkom}).
We  solve the
linearized equation for energy spectrum (for $t'=0$ and $\beta=50$)
and obtained the superconducting energy
gap on the Fermi surface (Fig.\ref{GAP}). The topology of the Fermi surface 
was defined as the zero-energy contour for the energy spectrum with all F's
Green functions equal to zero. It is clear that symmetry of the d-wave state
is not pure d$_{x^2-y^2}$ due to underlying AFM ordered states which lowered the
symmetry of d-SC. This also means that d-SC order could lower the symmetry
of the AFM Neel state and more general non-collinear magnetic state need to be
investigated.
Nevertheless the gap function has a maximum near the
$(\pi,0)$ and $(0, \pi)$ points and is almost zero near the $(\pi/2,\pi/2)$ point 
and the magnitude of the maximum superconducting gap is of the order of 15 meV
in a good agreement with experimental estimateess\cite{PGAP} and much lower then
the AFM gap for undoped case. 

In conclusion, we present the evidence for coexisting of antiferromagnetism
with the d-wave superconducting states within the cluster DMFT two dimensional
extended Hubbard model.
 
We thank Gabi Kotliar, Antoine Georges and Yurii Izyumov for helpful discussion.

\end{multicols}
\end{document}